\documentclass[useAMS,usenatbib]{mn2e}

\usepackage{color}
\usepackage{amsmath}

\usepackage{amsmath, mathtools}
\usepackage{amssymb}
\usepackage{wasysym}
\usepackage{graphicx} 
\usepackage{booktabs} 
\usepackage{pdfsync}
\usepackage{verbatim}
\usepackage{latexsym}
\usepackage{dsfont}

\title[Macro Dark Matter]{Macro Dark Matter}
\author[D. M. Jacobs, G. D. Starkman, and B. W. Lynn]
{David M. Jacobs$^{1,2}$\thanks{E-mail:
dm.jacobs@uct.ac.za},
Glenn D. Starkman$^{1}$\thanks{E-mail: glenn.starkman@case.edu}
and Bryan W. Lynn$^{1,3,4}$\thanks{E-mail: bryan.michael.lynn@cern.ch}
\\
$^{1}$ CERCA/Department of Physics/ISO, Case Western Reserve University, Cleveland, OH, 44106, USA\\
$^{2}$ Astrophysics, Cosmology and Gravity Centre and Department of Mathematics and Applied Mathematics,\\
~~University of Cape Town, Rondebosch 7701, Cape Town, South Africa\\
$^3$ University College London, London WC1E 6BT, UK\\
$^4$ Department of Physics, University of Wisconsin-Madison, Madison, WI, 53706, USA
}

\def\sec#1{\section{#1} }
\def\ssec#1{\subsection{#1} }
\def\sssec#1{\subsubsection{#1} }

\def\R{\\\\} 

\def\({\left(}
\def\){\right)}
\def\[{\left[}
\def\]{\right]}

\def\a{\alpha}

\def\f#1#2{\frac{#1}{#2}}
\def\g{\gamma}
\def\Ga{\Gamma}

\def\del{\nabla}
\def\tE{\widetilde{E}}

\def\l{\lambda}
\def\L{\Lambda}
\def\m{\mu}
\def\n{\nu}

\def\p{\pi}
\def\r{\rho}
\def\tr{\widetilde{r}}
\def\s{\sigma}
\def\t{\tau}
\def\th{\theta}

\def\cmsg{\text{~cm}^2~\text{g}^{-1}}
\def\gcmc{\text{g cm}^{-3}}

\def\eV{\text{eV}}
\def\keV{\text{keV}}
\def\MeV{\text{MeV}}
\def\GeV{\text{GeV}}

\def\cm{\text{cm}}
\def\km{\text{km}}

\def\kpc{\text{kpc}}

\def\bul{$\bullet$}

\def\<{\langle}
\def\>{\rangle}

\providecommand{\abs}[1]{\left\lvert#1\right\rvert}

\def\sX{\s_\text{\tiny X}}
\def\sXX{\s_\text{\tiny XX}}
\def\MX{M_\text{\tiny X}}
\def\QX{Q_\text{\tiny X}}
\def\RX{R_\text{\tiny X}}
\def\nX{n_\text{\tiny X}}
\def\rX{\r_\text{\tiny X}}
\def\vX{v_\text{\tiny X}}
\def\yX{y_\text{\tiny X}}
\def\sm{\sX/\MX}
\def\BX{B_\text{\tiny X}}

\begin{document}

\date{}

\pagerange{\pageref{firstpage}--\pageref{lastpage}} \pubyear{-}

\maketitle

\label{firstpage}

\begin{abstract}
Dark matter is a vital component of the current best model of our universe, $\L$CDM. There are leading candidates for what the dark matter could be (e.g. weakly-interacting massive particles, or axions), but no compelling observational or experimental evidence exists to support these particular candidates, nor any beyond-the-Standard-Model physics that might produce such candidates. This suggests that other dark matter candidates, including ones that might arise in the Standard Model, should receive increased attention. Here we consider a general class of dark matter candidates with characteristic masses and interaction cross-sections characterized in units of grams and $\cm^2$, respectively -- we therefore dub these macroscopic objects as \emph{Macros}. Such dark matter candidates could potentially be assembled out of Standard Model particles (quarks and leptons) in the early universe. A combination of Earth-based, astrophysical, and cosmological observations constrain a portion of the Macro parameter space. A large region of parameter space remains, most notably for nuclear-dense objects with masses in the range $55 - 10^{17}$ g and  $2\times10^{20} - 4\times10^{24}$ g, although the lower mass window is closed for Macros that destabilize ordinary matter.
\end{abstract}


\sec{Introduction}
Observations on all scales from galaxies up indicate that, unless General Relativity (GR) requires serious modification, we live in a universe whose energy content is dominated by substances that differ from our everyday experience. The best dynamical model of the evolution of the Universe and its contents consistent with GR is $\L$CDM, which describes a universe whose energy density is dominated on the largest scales by a cosmological constant ($\Omega_\L\simeq 0.7$) and a non-relativistic matter component ($\Omega_\text{\tiny m}\simeq 0.3$).  Observations of galaxies and clusters suggest that, assuming the correctness of Newton's inverse square law of gravity, they are heavily dominated by non-relativistic matter that cannot be accounted for in any census of their ordinary baryonic matter.
At the same time, the observed primordial abundances of the light elements tell us that the fraction of the cosmological energy density that is due to baryons, $\Omega_\text{\tiny b}\lesssim0.05 \ll \Omega_\text{\tiny m}$ (\cite{McGaugh:2007fj,Ade:2013zuv}). The remaining fraction of the non-relativistic matter, $\Omega_X\equiv \Omega_\text{\tiny m}-\Omega_\text{\tiny b}$, must be some kind of weakly-interacting matter that, apparently, is not a particle of the Standard Model. 

The leading dark matter candidates are supersymmetric thermal relics, a class of stable weakly-interacting massive particles (WIMPs) that arise in certain theories of low-energy supersymmetry; however searches for supersymmetry at the LHC (\cite{Robichaud-Veronneau:2013lma, CMS:2013gsa}) have so far failed to discover anything. Likewise, direct detection experiments (e.g. \cite{Akerib:2013tjd, Aprile:2012nq, Agnese:2013rvf}) have yet to make any conclusive detection of conventional WIMPs. 

We have few clues about the nature of the dark matter, except that, based on observations, it must satisfy a series of negative requirements: it shouldn't ruin the success of big bang nucleosynthesis (BBN) nor the physics of the cosmic microwave background (CMB); large scale structure must be allowed to grow to form galaxies and clusters, and the dark matter must remain undetected in any of the direct or indirection measurements.  In fact, dark matter might only interact gravitationally  -- anything more than this would simply be the result of nature being kind to us.  As usual, we proceed under this more optimistic assumption.


Let us take a step back to consider that there are two possibilities about the nature of the dark mater: (I) it is intrinsically weakly interacting, or (II) it is \emph{effectively} weakly interacting because it is massive and hence has a much lower number density. Dark matter-baryon interaction rates go as $\sim\nX \sX v$, the product of the dark matter number density, the interaction cross-section, and a characteristic velocity. Since $\nX=\rX\MX^{-1}$ and $\rX$ is fixed for any dark matter scenario, the event rate is proportional to $\sm$, which we call the \emph{reduced cross-section}. Conventionally, dark matter is dark because $\sX$ is small; this is possibility (I). But it can equally be dark if $\MX$ is very large; this is case (II) and is what we are interested in this work. An interesting possibility that case (II) allows for is that the dark matter might still be accounted for within the Standard Model.

For example, given that the local dark matter density is measured to be about $7\times10^{-25}\,\gcmc$ (\cite{Beringer:1900zz}) and the characteristic velocity of the dark matter is presumably about $10^{-3}\,c$, dark matter objects with masses on the order of $10^{18}g$ would hit the Earth approximately once every billion years. At lower masses the frequency would be higher, but the nature of the impact matters greatly as to whether or not some signal is observable by humans or if some historical record was left to be discovered.


Of course, this basic notion is not entirely new. Consider, for example, a proposal by \cite{Witten:1984rs} wherein the QCD phase transition in the early universe resulted in an abundance of baryons alongside macroscopically sized/massed ``nuggets" of quark matter with an approximate nuclear density of a few $\times~10^{14}~\gcmc$. Estimates in that work suggested the mass of a typical nugget, which is posited to be the dark matter, could be $10^{9} - 10^{18}$ g. At this range of masses, the expected rate of collision between such dark matter and the \emph{entire} Earth is at most once per year; clearly, underground detection experiments will have nothing to say about this possibility.  Though the so-called WIMP miracle doesn't exist, what is highly appealing in such a scenario is that little to no new physics is invoked to explain the origin of dark matter and, as a corollary, it offers a natural explanation as to why  $\Omega_\text{\tiny m} \sim \Omega_\text{\tiny b}$. There is an abundance of models directly (or indirectly) associated with this type of idea, for example: 
nuclearites (\cite{DeRujula:1984ig}), 
strangelets (\cite{Farhi:1984qu}), 
strange baryon Q-Balls (\cite{Lynn:1989xb}), 
baryonic colour superconductors  (\cite{Zhitnitsky:2002nr,Zhitnitsky:2002qa}), 
compact composite objects (CCO's) (\cite{Zhitnitsky:2006vt}), 
strange chiral liquids drops (\cite{Lynn:2010uh}),
and Compact Ultradense Objects (CUDOs) (\cite{Labun:2011wn}). 

There are also primordial black holes (PBH) (\cite{Carr:1974nx}), for which there have been extensive efforts to constrain as dark matter candidates (see e.g. \cite{Carr:2009jm, Capela:2012jz, Capela:2013yf, Pani:2014rca, Belotsky:2014kca}), non-Standard Model candidates associated with new hidden symmetries (e.g. \cite{Kusenko:1997si, Khlopov:2013ava, Murayama:2009nj, Derevianko:2013oaa, Stadnik:2014cea}), and any other such objects not yet hypothesized that could make up some or all of the dark matter.


While specific theories have their own appeal, we find it prudent to try to understand the phenomenology of a general class of models in which the dark matter interacts with itself and normal matter strongly; in other words, its interaction probability is determined predominantly by geometry and kinematics. Existing constraints (as summarized in \cite{Mack:2007xj}) on strongly-interacting dark matter cover large regions of parameter space extending to masses of about $10^{17}$GeV, prompting us to consider massive candidates with radius, $\RX$ much larger than any microscopic length scale, e.g. the electron's Compton wavelength or the Bohr radius. We can then ignore any quantum-mechanical aspects of scattering, and any short-range interaction will simply be encoded in the dark matter's geometric\footnote{Gravitational focusing is negligible since the escape velocity is typically very small compared to the characteristic velocities of order $10^{-3}c$.} cross-section, $\sX=\p \RX^2$. It may also interact electromagnetically -- we therefore consider dark matter objects with a charge $\QX$. Generally, for these types of models the effective cross-section and mass are best quoted in $\cm^2$ and g, respectively. We call this class of ``macroscopic" dark matter \emph{macro dark matter} and refer to the objects as \emph{Macros}. 

Assuming Macros are formed by some post-inflationary causal process, they have a maximum mass determined by the amount of dark matter within the causal horizon at the time of formation, $M_{H,\text{dark}}$, given by 
\begin{align}
M_{H,\text{dark}}&\simeq\f{4\p}{3}\r_{\text{\tiny X}}\(T_\star\)L_H^3\notag\\
&\sim 10^{35}~\text{g} \(\f{10^9~\text{K}}{T_\star}\)^3\,,
\end{align}
where $T_\star$ is the formation temperature. Although the dark matter could have formed as late as matter-radiation equality if BBN is not disturbed, we shall assume the formation processes finished before BBN, \emph{therefore it may be baryonic or non-baryonic in nature.} Assuming $T_\star \gtrsim $ few $ \times 10^9$ K means we consider only $\MX \lesssim 2\times 10^{34}\text{g}= 10 M_\odot$.




The impact rate of an isotropic flux of Macros hitting a convex target object is\footnote{One factor of $1/2$ comes from the average of the component of the velocity vector normal to the surface, i.e. from the angular average of $\cos \th$, while the other factor of 1/2 is included to avoid counting up-going impactors.}
\begin{equation}\label{impact_rate}
\Gamma=\f{1}{4} \nX \vX A_\text{\tiny T} f_\text{G}\,,
\end{equation}
where $A_\text{\tiny T}$ is the target area, $\vX$ is the average Macro velocity, and $f_\text{G}=(1+  \f{      v_\text{esc}^2  }{   \vX^2     }    )$ is the gravitational focusing factor.  Altogether, the total impact rate is
\begin{align}\label{impact_rate_2}
\Gamma=2.7 \times 10^5&\,\text{s}^{-1} \(\f{1 \text{g}}{\MX}\) \(\f{\vX}{250~\text{km s$^{-1}$}}\)\(\f{R_\text{\tiny T}}{R_\odot}\)^2\notag\\
&\times\(1+6.2  \f{R_\odot}{R_\text{\tiny T}}\(\f{250~\text{km s$^{-1}$}}{\vX}\)^2 \f{M_\text{\tiny T}}{M_\odot}     \) f_\r \,,
\end{align}
where $M_\text{\tiny T}$ and $R_\text{\tiny T}$ are the mass and radius of the target, $R_\odot=7\times 10^5\,\km, M_\odot=2\times 10^{33}$ g, we have used $\MX\nX=\rX=7.0 \times 10^{-25}\,\gcmc$ as the local dark matter density (\cite{Beringer:1900zz}), and defined $f_\r$ as a density enhancement factor that is equal to unity in the solar neighborhood. In Table \ref{table:hitrates} we give the expected impact rates for various astrophysical objects.
\begin{table}
\centering 
\begin{tabular}{c c c} 
\hline\hline 
Target& $\Gamma$ [$\MX^{-1} \text{g~}  \text{s}^{-1}$]   & $\Gamma$ [$\MX^{-1} \text{g~}  \text{yr}^{-1}$]   \\ [0.5ex] 
\hline 
NS & $ 24$ & $7.5 \times 10^8$   \\ 
WD & $ 2.4\times 10^3$ & $7.5 \times 10^{10}$   \\ 
$\odot$ & $ 1.9 \times 10^6$& $6.1 \times 10^{13}$  \\  
$\oplus$ & $ 22$& $6.9 \times 10^{8}$ \\  
\leftmoon & $ 1.6$ & $5.0 \times 10^{7}$ \\ [1ex] 
\hline 
\end{tabular}
\caption{Expected Macro impact rates for a neutron star, white dwarf, the Sun, the Earth, and the Moon. We have taken $\vX=250~\text{km s$^{-1}$}$, $R_\text{NS}=10$ km, $R_\text{WD}=10^3$ km, $f_\r=1$, and $M_\text{NS}=M_\text{WD}=M_\odot$. For example, if $\MX=1$ g then there would be about 1 impact per $\km^2$ per year on the Earth. } 
\label{table:hitrates} 
\end{table}

Throughout this work we assume for simplicity that the Macros have a single mass. In Section \ref{model-indp_constraints} we give model-independent constraints, including those extant in or extracted from the published literature and applied to Macros. In Section \ref{model-dep_constraints} we give constraints that depend on specific Macro properties, such as electromagnetic charge. In Section \ref{ideas} we report on considerations that do not appear to provide any useful constraints, and we make our concluding remarks in Section \ref{conclusions}. Throughout, we set $\hbar=k_\text{\tiny B}=c=1$.

\sec{Model-Independent Constraints}\label{model-indp_constraints}

\ssec{Constraints at low masses}
There are a variety of underground and space-borne detectors that   have been used to put constraints on a large range of strongly-interaction dark matter parameters below $10^{17}\GeV$; this is not the focus of the current work, but the list of model-independent constraints may be found in \cite{Mack:2007xj}. However, since the constraints obtained from the Skylab space station overlaps somewhat with our work here, we briefly summarize those results. 

The Lexan (plastic) track detectors inside a wall of the Skylab space station (\cite{shirk1978charge}) were used to probe the nature of cosmic rays and also were used to rule out a region of parameter space of strongly-interacting dark matter. The details of the dark matter constraints are discussed in \cite{Starkman:1990nj} and we simply summarize the salient points below.

For elastically-scattering Macros, the energy loss rate of $\r^{-1} dE/dx$ in the Lexan must have exceeded a minimum $400\,\MeV \cmsg$  beyond which enough damage was done to the Lexan that an etchable track would have been seen. Since $\r^{-1} dE/dx\simeq\sX v^2$, if $\vX\simeq 250\,$km s$^{-1}$ then the requirement is $\sX\geq10^{-18}~\cm^2$; this determines the bottom edge of the Skylab-constrained region of the $\sX-\MX$ parameter space. The Macro also must penetrate a minimum distance of $x_\text{min}\simeq 0.25$ cm while maintaining the above energy loss criterion for it to be visible. The velocity of such Macros will decrease in the material according to
\begin{equation}
v(x_\text{min})=v_0 e^{-\r x_\text{min} \sm  }\,.
\end{equation}
As in \cite{Starkman:1990nj}, we assume $x_\text{min}=0.25 ~\cm$ so that $\r x_\text{min}\sim 0.25$ g cm$^{-2}$ which implies $\sm\lesssim 3\cmsg$ to use the Skylab constraints. For inelastically-scattering Macros, the requirement for an etchable track is that the hole cleared out in the Lexan could have been large enough that chemical reagents could have entered the hole during the etching process. This is plausible for hole radii larger than a few angstroms, or $\sX\gtrsim 10^{-15}\cm^2$.

The total exposure of the Skylab experiment was on the order of  $2~\text{m}^2  ~\text{yr}~\text{sr}$. Given a dark matter flux of $(4\pi)^{-1} \nX \vX \simeq 2\times10^{17}(\GeV\,\MX^{-1})~\text{m}^{-2}  ~\text{yr}^{-1}~\text{sr}^{-1}$, the Skylab results rule out macro dark matter satisfying the above criteria for masses below about $10^{17}\GeV\simeq2\times 10^{-7}$ g at greater than $95$ per cent confidence.

\ssec{The Large-Scale Universe}

\sssec{Constraints from self-interacting dark matter}

For a given dark-matter density ($\rX$) and mean free path of $L_\text{free}$ due to Macro-Macro collision the cross-section ($\sXX$) will obey the relation
\begin{align}
\f{\sXX}{\MX}&=\f{1}{\rX}\f{1}{L_\text{free}}\notag\\
&\simeq 0.5 \(\f{7.0 \times 10^{-25}\gcmc}{\rX}\) \(\f{1 ~\text{Mpc}}{L_\text{free}}\) \cmsg\,.
\end{align}

Self-interacting dark matter was proposed to solve inconsistencies between CDM predictions and observations of  structures on scales below a few Mpc, including the cusp-core and missing-satellite problems  (\cite{Spergel:1999mh}). These inquiries have prompted several investigations of the strength of possible dark matter self-interactions.  Simulations of galaxy cluster 1E 0657-56 (a.k.a. the bullet cluster) accounting for the self-interaction of dark matter result in an offset between the bullet sub-cluster mass peak and galactic centroid;  the absence of this observation in the actual cluster therefore provides a limit on $\sm$ (\cite{Randall:2007ph}).  Comparisons were also made between simulations with self-interacting dark matter and the observed density profiles and substructure counts of other observed clusters, low-surface brightness spiral- and dwarf-spheroidal galaxies in \cite{Rocha:2012jg}. Conservatively, both of the above constraints are approximately $\sXX \,\MX^{-1} \lesssim1 \, \cmsg$. From simple geometric considerations, the self-interaction cross-section of two Macro's is related to their geometric cross-section by a factor of 4, i.e. $\sXX=4\,\sX$; therefore the constraints are $\sm \lesssim 0.3 ~\cmsg$ for elastically-scattering Macros.

Conservatively, the constraint for inelastically-scattering Macros should be at least this strong because in such a scenario Macros should have never collided on average. However, requiring a mean free path greater than the distance traveled in the age of the universe, $\vX \Delta t\simeq10^{-3}c\, H_0^{-1}\simeq 3$ Mpc would impose the constraint $\sm\lesssim 0.04~\cmsg$. Stronger limits may still be obtainable on the inelastic case based on observations of dense galactic regions.


\sssec{Constraints from Macro-baryon interaction}


Virialized particles in a gravitational potential reach similar velocities; however when particles of different species collide elastically there is a preferential energy transfer from massive to less massive particles, i.e. energy is transferred to drive the system toward thermal equilibrium.  For this reason, Macros would provide a source of heat for gas in astrophysical systems, such as clusters. As was illuminated in \cite{Chuzhoy:2004bc} such gas heating, which would occur at a rate proportional to $\sm$, can offset radiative cooling in the cores of clusters. To avoid conflict with observations, namely that gas temperatures do not increase toward the centers of clusters, they found an upper bound that corresponds to $\sm<10^{-25}$ cm$^2$ $m_p^{-1}$$=0.06 \cmsg$ for elastic Macro-baryon interactions.

Dark matter-baryon interactions would also result in a drag force between the two fluids at early times and would act to dampen fluctuations at small length scales that would result in a suppression of the growth of massive structures, as was investigated by \cite{Boehm:2000gq}, \cite{Chen:2002yh}, and \cite{Boehm:2004th}.

 
In order to justify the application of this formalism to Macros we must demonstrate that the low number density of the Macros does not ruin the dark matter fluid approximation. First, we must ensure that the physical volumes considered are much larger than $\nX^{-1}$. Given an average cosmological dark matter density of $\simeq 2\times10^{-30}\gcmc$ and a maximum Macro candidate mass of around $10^{34}$ g requires us to consider comoving volumes $\gg 1~\kpc^3$, which is true of the cosmological probes considered here. 
 
Secondly, we must ensure that the diffusion time, $\t_\text{diff}$, for a baryon to cross the average Macro separation, $L_\text{DM}=\nX^{-1/3}$, must be short compared to the relevant Hubble time. The mean free path of a baryon is determined by its interaction with other baryons and background photons. For baryon-baryon collisions it is $\l_{bb}\simeq (n_b \s_{bb})^{-1}$ where the characteristic cross section, $\s_{bb}\sim \a^2/p_b^2\sim \a^2/(m_b T)$. 
For baryon-photon collisions, since we're interested in the era where the temperature is much smaller than the baryon mass, the photon momentum transfers to the baryons are relatively small. They are also random walks in momentum space, therefore the effective mean free path from baryon-photon collisions for randomizing the baryon momentum is $\l_{b\g}\simeq \(n_\g \s_\text{p-$\g$}v_b\)^{-1}$, where $\s_\text{p-$\g$}\simeq\s_\text{\tiny T}\(m_e/m_p\)^2$.  Given that the dominant contribution to the collisional damping comes from times close to matter-radiation equality at $z\simeq 3000$ (see e.g \cite{Boehm:2000gq}), it is the baryon-baryon collisions that are most significant here.  
The baryon diffusion length is $\l_D\simeq\l_\text{MFP} \sqrt{N}$, where $N=\t_\text{diff}\,v_b/\l_\text{MFP}$ is the effective number of collisions. It follows that a baryon will diffuse over the distance $L_\text{DM}$ in a time
\begin{equation}
\t_\text{diff,DM}=\f{L_\text{DM}^2}{\l_\text{MFP}v_b}
\end{equation}
which must be small compared to the Hubble time, $H^{-1}$. In other words
\begin{equation}
 \f{ n_b\a^2}{\nX^{2/3}m_b^{1/2} T^{3/2}}   H  \ll 1\,. 
\end{equation}
In terms of the Macro mass and redshift, this inequality may be written as approximately
\begin{equation}
\(\f{\MX}{1\,\text{g}}\)^{2/3}\(1+z\)^{3/2}\ll 10^{29}\,.
\end{equation}
Even for the largest Macro candidates of about $10^{34}~\text{g}$, this inequality is at least marginally obeyed for $z\lesssim10^4$, therefore the Macro fluid approximation is valid for this analysis.
 
In \cite{Dvorkin:2013cea}, the effects of velocity-dependent dark matter-baryon interactions on the CMB and Lyman-$\a$ power spectrum analysis were considered, and the lack of power spectrum suppression was used to infer a bound on the dark matter-baryon interaction. 
To constrain Macro properties, we borrow those velocity-independent results to place the bound of
\begin{equation}
\f{\sX}{\MX} < 
3.3 \times 10^{-3} \cmsg
\end{equation}
on elastically-scattering Macros at 95 per cent confidence.

\sssec{Constraints from Macro-photon interaction}

Macro-photon coupling would result in a similar type of collisional damping found in the Macro-baryon case as considered in \cite{Boehm:2000gq}, \cite{Boehm:2001hm}, and \cite{Boehm:2004th}. 

As in the last section, we must ensure that the diffusion time, $\t_\text{diff}$, for a photon to cross the average Macro separation, $L_\text{DM}=\nX^{-1/3}$, is short compared to the Hubble time. The diffusion length, $\l_D=\l_\text{MFP} \sqrt{N}$, where $\l_\text{MFP}=\(n_e \s_\text{T}\)^{-1}$ and $N=\t_\text{diff}/\l_\text{MFP}$ is the number of collisions. A photon will diffuse over the distance $L_\text{DM}$ in a time
\begin{equation}
\t_\text{diff,DM}=\f{L_\text{DM}^2}{\l_\text{MFP}}
\end{equation}
which must be small compared to the Hubble time, $H^{-1}$, or
\begin{equation}
n_e \s_T \nX^{-2/3} H  \ll 1\,.
\end{equation}
In terms of the Macro mass and redshift, this inequality may be written as approximately
\begin{equation}
\(\f{\MX}{1\,\text{g}}\)^{2/3}\(1+z\)^3\ll 10^{41}\,.
\end{equation}
For the largest Macro candidates of about $10^{34}~\text{g}$, $\t_\text{diff,DM}<H^{-1}$ remains true for $z\lesssim10^6$. Therefore, as in the case of Macro-baryon interaction, this analysis is justified.

The collisional damping scale can be computed analytically and, based on the existence of structures above a certain mass (corresponding to density fluctuations at a given length scale), bounds on possible dark matter-Standard Model interactions can be derived (\cite{Boehm:2000gq, Boehm:2001hm, Boehm:2004th}). In \cite{Wilkinson:2013kia} the full Boltzmann formalism was utilized to numerically obtain constraints on elastic dark matter-photon interactions using CMB data from the Planck satellite. A bound of
\begin{equation}\label{CMB_bound}
\sm < 8\times10^{-31} \text{cm}^2/\GeV \simeq4.5\times10^{-7}\cmsg
\end{equation}
on the elastic scattering cross section between dark matter and photons was determined. This bound, originally developed for particle-type dark matter, apparently also applies to the elastic scattering of photons from Macros.

\begin{figure}
  \begin{center}
    \includegraphics[scale=.45]{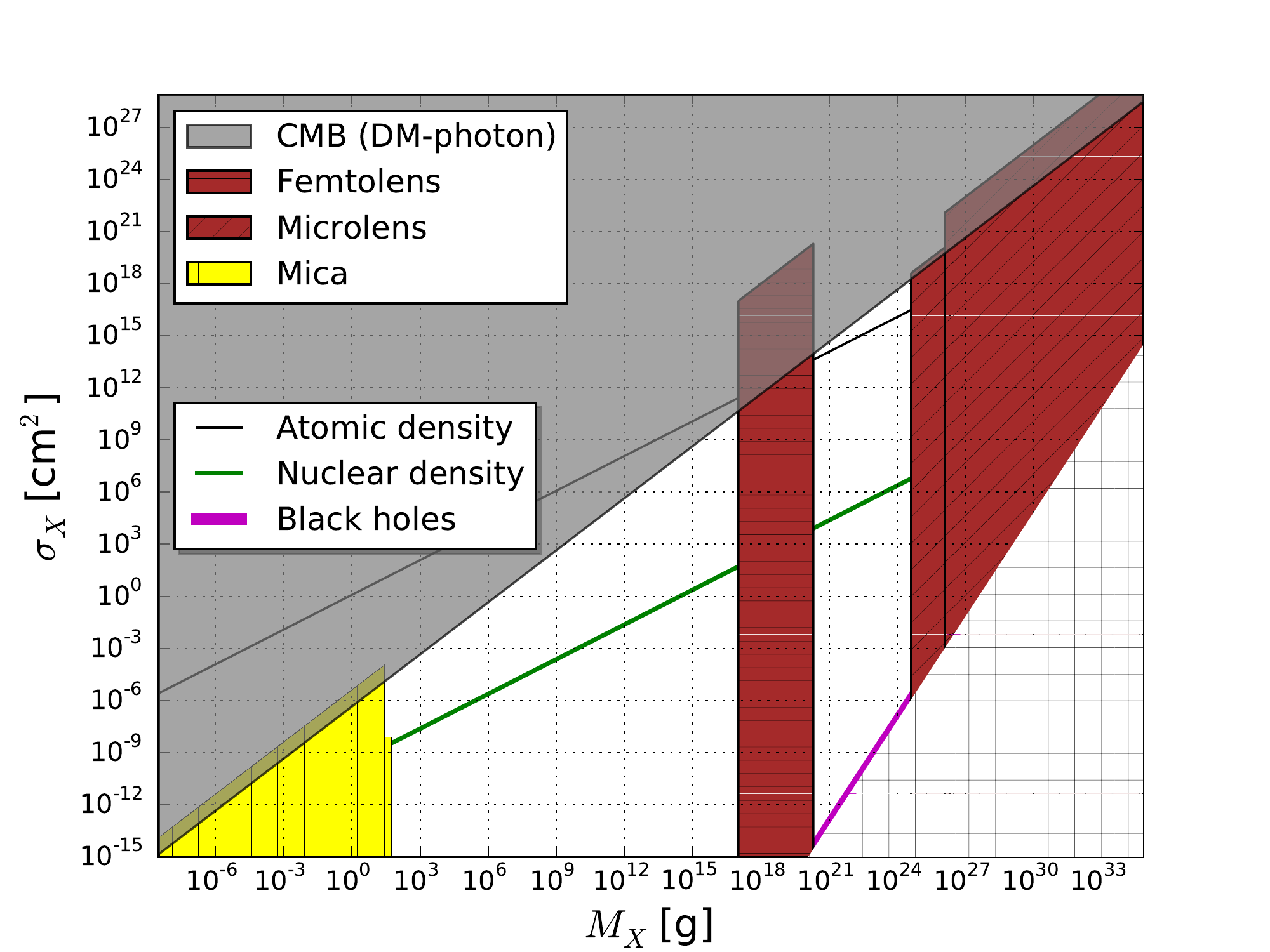}
  \end{center}
\caption{Constraints on the Macro cross section and mass (assuming the Macros have a single mass), including the Macro-photon constraints and applicable for both elastically- and inelastically-scattering candidates. In red are the femto- and micro-lensing constraints, while in grey are the CMB-inferred constraints (see text for references). The black and green lines correspond to objects of constant density $1~\gcmc$ and $3.6 \times 10^{14}~\gcmc$, respectively. Black hole candidates lie on the magenta line, however, these may be ruled out for other reasons (see \citet{Capela:2012jz, Capela:2013yf, Pani:2014rca, Defillon:2014wla}); objects within the hatched region in the bottom-right corner should not exist as they would simply be denser than black holes of the same mass.}
\label{model_independent}
\end{figure}

If Macros are perfect absorbers, i.e.  if they are blackbodies, they would also emit  a thermal spectrum given by the local temperature (corresponding to the monopole of the distribution function) and would do so isotropically, or at least their emission would be statistically isotropic. It may be concluded that the collision term appearing in the Boltzmann equation therefore must be exactly the same as for the elastic case.  In a generic scenario, Macros that interact with photons would have some reflection coefficient (albedo), but since both reflection and absorption/re-emission apparently have the same effect, it is apparently \emph{independent} of the albedo. Therefore the CMB bound \eqref{CMB_bound} on Macro-photon interactions applies for  \emph{any} reflectivity of the Macros so long as they maintain thermal equilibrium with the plasma. In Figure \ref{model_independent} we include these Macro-photon derived constraints in our model-independent constraints.
 

\ssec{Ancient Mica}

If the Macros have a low enough mass, their number density (and hence flux) would be high enough to have plausibly left a historical record on Earth. If they have a low enough $\sm$ so that they would have penetrated deep ($\sim$ a few km) into the Earth's crust, a record would have been left in ancient muscovite mica. Searches for grand-unified-theory magnetic monopoles (\cite{Price:1983ax, Price:1986ky}) sought to detect lattice defects left in ancient mica, detectable through chemical etching techniques (see e.g. \cite{fleischer1998tracks}).  These same techniques were applied to put limits on the astrophysical flux of so-called nuclearites, in the range $10^{-16} - 10^2$g  (\cite{DeRujula:1984ig, Price:1988ge}).  Here we apply similar arguments to place constraints on a region of Macro parameter space.

\begin{figure}
  \begin{center}
    \includegraphics[scale=.45]{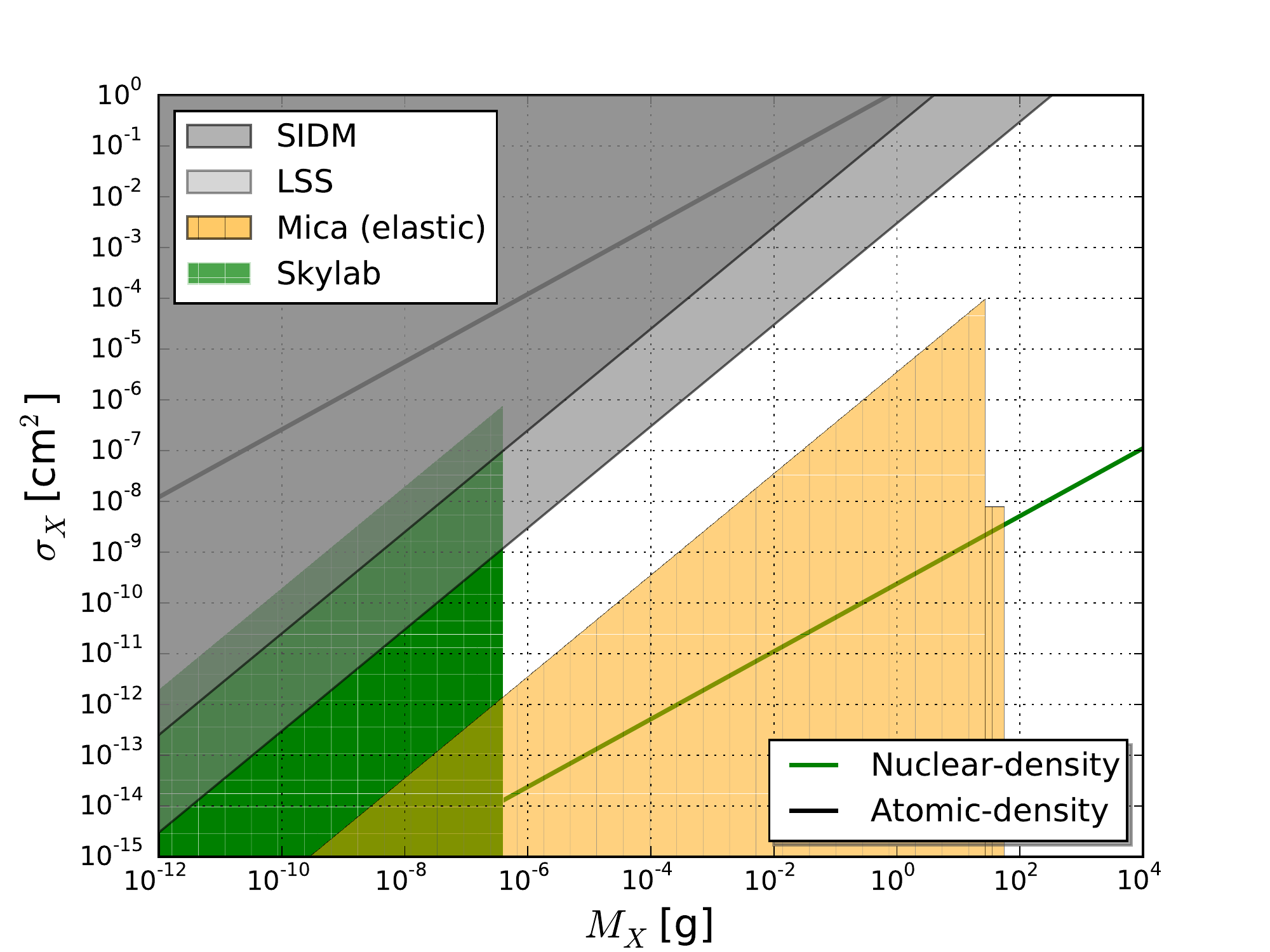}
  \end{center}
\caption{Constraints on Macro cross section and mass, assuming the Macros have a single mass, inferred from \emph{elastic} scattering with baryons and other Macros. In dark grey are those inferred from comparison of observed to numerically-simulated galaxies and clusters with self-interacting dark matter (\citet{Randall:2007ph, Rocha:2012jg}); overlapping, in light grey are the large-scale structure constraints taken from \citet{Dvorkin:2013cea}; the orange region correspond to the mica constraints (\citet{Price:1986ky}); the green region is the Skylab constraint (\citet{shirk1978charge, Starkman:1990nj}). The black and green lines correspond to objects of constant density $1~\gcmc$ and $3.6 \times 10^{14}~\gcmc$, respectively.}
\label{elastic_plot}
\end{figure}

\begin{figure}
  \begin{center}
    \includegraphics[scale=.45]{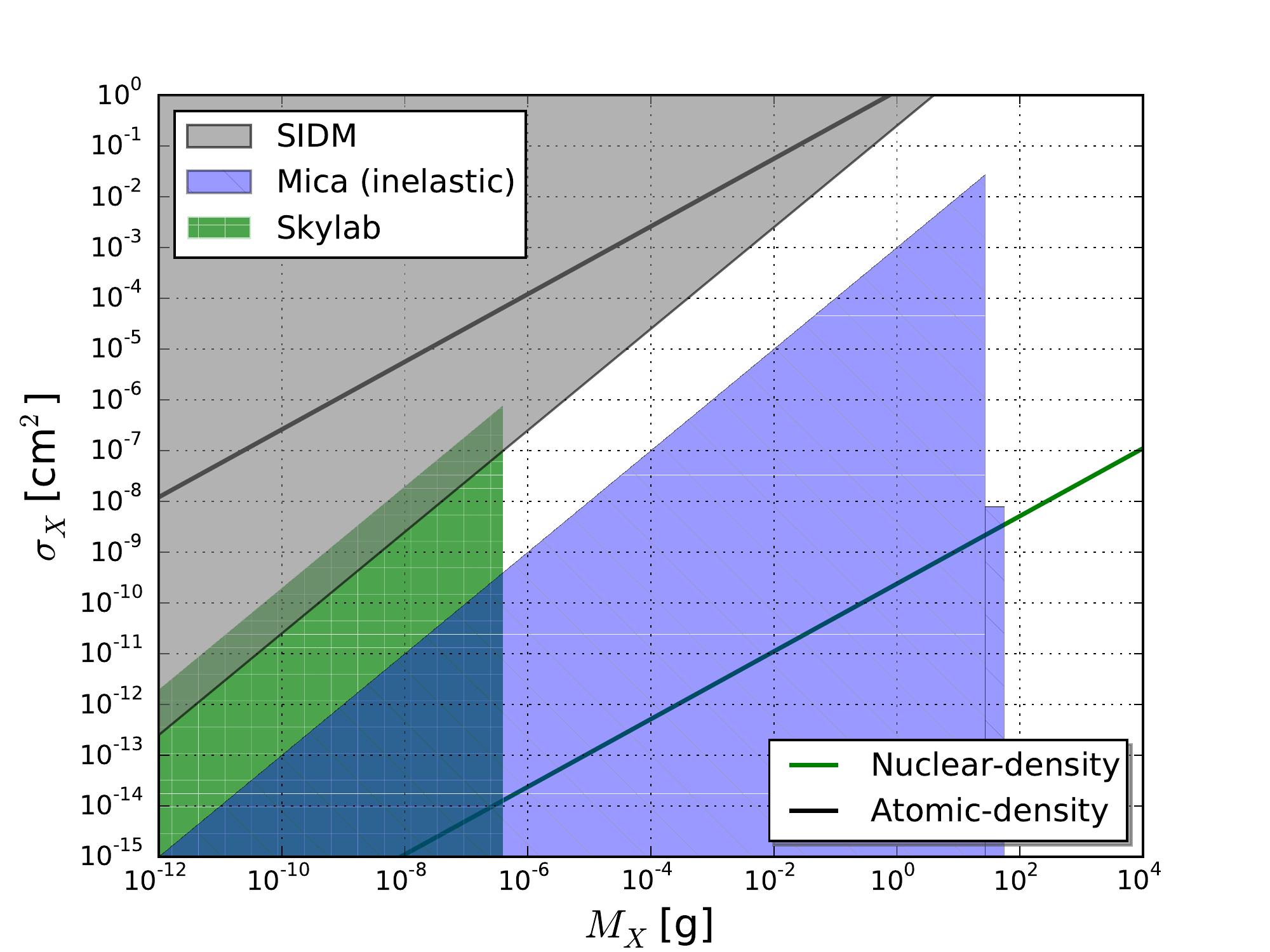}
  \end{center}
\caption{Constraints on Macro cross section and mass, assuming the Macros have a single mass, inferred from \emph{inelastic} scattering with baryons. Here, the grey region is the constraint applied from self-interacting dark matter (which the Macro-baryon interaction should not exceed) and the blue region corresponds to the inelastic mica constraints.}
\label{inelastic_plot}
\end{figure}

The constraining power of mica is determined by a few factors. No significant detection was found in the samples considered in \cite{Price:1986ky} which have an approximate age of $500$ Myr, a combined total surface area of about $2400~\cm^2$, and was buried approximately 3 km underground. Using \eqref{impact_rate}, an expected $\sim165$g of dark matter should have passed through the sample\footnote{This assumes that Macros pass undeterred straight through the Earth. For $\sm \gtrsim \(\r_\oplus R_\oplus\)^{-1}\simeq 3\times 10^{-10}\,\cmsg$ the value of $\l$ is decreased, and our Macro bound weakened, by a factor of 2.}, or an average number of mica passages, $\l=165\,\text{g}/\MX$. Since a Macro impact is a random (Poisson) process, the probability of $n$ passages, $P(n)$, follows a Poisson distribution:
\begin{equation}
P(n)=\f{\l^n}{n!}e^{-\l}\,.
\end{equation}
Given the null observation, the value $\l\gtrsim3$ may be ruled out at 95 per cent confidence, or translating this to a bound on $\MX$,
\begin{equation}
\MX \gtrsim 55\, \text{g}\,,
\end{equation}
where the mica constraints are applicable; however the bound is weakened to about $28$ g for $\sX\gtrsim10^{-8}\, \cm^2$, and this is accounted for in Figures \ref{elastic_plot} and \ref{inelastic_plot}. At a characteristic nuclear density of $3.6\times 10^{14}\,\gcmc$, we infer $\RX\gtrsim 3\times 10^{-5}\text{~cm}$ if the Macro would admit a nuclear/QCD description.

There are also detection thresholds in velocity and energy deposition for elastic scattering. The Macro must have had a velocity greater than $2\times 10^{-5} c$ upon reaching the buried mica\footnote{This is due to kinematic reasons (\cite{Price:1983ax}); it corresponds to a nuclear energy transfer of about $0.2 A~\eV $, where $A$ is the atomic number.}. For elastically-scattering Macros, the velocity at an average projected depth in the crust below the Earth's surface, $\<\r L\>$, is approximately
\begin{equation}
v(L)=v_0 e^{-\<\r L\> \sm }\,,
\end{equation}
where we use $v_0\equiv\<v^2\>^{1/2}$ as the initial velocity\footnote{Technically, the Macros would have velocities are drawn from a distribution, but the $\sm$ requirement is only log-sensitive to the velocity, hence, we will ignore the tails of the distribution.}. Taking $v_0=250$ km s$^{-1}$ and $\<\r L\>=10^6$ g cm$^{-2}$, the imposition of $v(L)> 2\times 10^{-5}c$ requires
\begin{equation}
\f{\sX}{\MX} \lesssim 3.6\times 10^{-6} \cmsg\,
\end{equation}
to obtain a constraint\footnote{In the crust $2.7 \lesssim \r \lesssim 3.0 \,\gcmc$. We conservatively use $\r=3~ \gcmc$ and a mica depth of $L=3.5$ km to estimate $\<\r L\>$.}.

In order to have left an etchable track, there is also a minimum nuclear component of stopping power $S_n\equiv \r^{-1} dE/dx\gtrsim 2.4~ \GeV/\text{g~} \cm^2$ (\cite{Price:1986ky}).  For an elastically interacting Macro, $S_n \simeq \sX v\(L\)^2$, so constraints require 
\begin{equation}
\sX \gtrsim 6 \times 10^{-18}~\cm^2 ~\(\f{250~\text{km s}^{-1}}{v(L)}\)^2
\end{equation}
to be satisfied. For masses above approximately $10^{-10}$g we expect a track to have been left for $\sX\gtrsim 6 \times 10^{-18}~\cm^2$. However at lower masses this inequality is less accurate since objects with smaller masses would have velocities that are more affected by their passage through the Earth's crust.

For inelastic collisions, crustal material is accreted onto the Macro. Since our aim is to make a conservative, model-independent constraint we require that the Macro's velocity would have remained above a critical value, $v_c=\sqrt{\varepsilon/\r}\simeq 0.2$ km s$^{-1}$, below which the energy loss from inter-molecular bond-breaking would have rapidly brought the Macro to rest; we have used a structural energy density of $\varepsilon \simeq 10^9$ erg cm$^{-3}$ as in \cite{DeRujula:1984ig}. For inelastic collisions, $v(L)=v_0\(1+ \<\r L\>\sm\)^{-1}$, therefore $v(L)>v_c$ implies $\sm<v_0/v_c (\r L)^{-1}\simeq10^{-3}\cmsg$. The requirement for an etchable track is that the burrowed hole in the mica sample would have been large enough that hydrofluoric acid would have entered it during the etching process. This is plausible for hole radii larger than a few angstroms, so $\sX\gtrsim 10^{-15}\,\cm^2$ should suffice.
 There could also be some charge-dependent enhancement to this process, however this is likely to be very material-dependent.

\ssec{Gravitational Lensing}

For an intervening mass between a light source and an observer, two images of the source can be observed, in principle.  Below we give a brief review of the physics of lensing before proceeding to discuss how the phenomenon (or the lack of its observation) may be used to place constraints on the abundance of Macros.

We use the canonical variables to describe the the observer-lens and lens-source distances as $D_\text{OL}$ and $D_\text{LS}$, respectively. The total observer-source distance is the sum of the two, denoted $D_\text{OS}$. Working in the lens plane with lens at the origin of coordinates, the line directly between the source and the observer is projected onto this plane at a distance, $r_0$, from the lens. Generally, the lens causes the observer to see two images\footnote{If $r_0\simeq0$ then a so-called Einstein ring would be observed.}; the length scale characteristic of the impact parameter of a wave packet as it travels past the lensing mass is given by the Einstein radius,
\begin{equation}
R_\text{\tiny E}=\sqrt{4GM\f{D_\text{OL}D_\text{LS}}{D_\text{OS}}}\,.
\end{equation}

Two electromagnetic waves passing in the vicinity of the lens with characteristic photon energy, $E$, and arriving at the location of the observer have amplitudes that add and interfere, resulting in an intensity (or flux) amplification given by
\begin{align}
A
&= \f{2+u^2 + 2\cos{\Delta\phi}}{u\sqrt{4+u^2}}\,,
\end{align}
where $u\equiv \f{r_0}{R_\text{\tiny E}}$, and the phase difference $\Delta \phi=E\Delta r$ between the two waves is determined by both the energy of the photons and difference in their path lengths, $\Delta r$.

One way to detect the presence of an intervening lens is to look for a modulation of an observed photon flux as a function of $E$ -- this is the basic idea behind femtolensing (\cite{gould1992femtolensing}). When considering  sources for which $r_0$ changes substantially in an observing time, the amplification would vary as a function of time -- this is the basis of microlensing (\cite{Paczynski:1985jf}). Historically, the above techniques have been used to place constraints on the abundance of primordial black holes, brown dwarfs, or other compact astrophysical objects comprising a fraction of the dark matter density. To constrain the abundance of Macros, we will (conservatively) insist that the Einstein lensing angle, $\th_E=\sqrt{4G\MX/d}$ exceed the angular size of the Macro, $\th_X=\RX/d$, for a characteristic distance, $d$. This amounts to the requirement
\begin{equation}\label{lensing_reqt}
\sm \lesssim 3 \times \(\f{d}{\text{Gpc}}\) \cmsg\,.
\end{equation}

\sssec{Femtolensing}

Femtolensing refers to gravitational lensing where the angular separation between two lensed images from the same source is of order $10^{-15}$ arcseconds. At such scales, images cannot be resolved, however an interference pattern in the energy spectrum of background sources would be observable. For a gamma-ray burst (GRB), for example, the magnification is energy-dependent and this results in an intensity pattern in the energy spectrum (\cite{gould1992femtolensing, Barnacka:2012bm}).

The lensing probability is determined by the optical depth to the source, $\t$ and a ``lensing cross section" (\cite{Barnacka:2012bm}). First, the optical depth may be calculated by noting that the cross section for ``strong" lensing events is characteristically given by (see e.g. \cite{fukugita1992statistical})
\begin{align}
\s &= \p R_\text{\tiny E}^2\notag\\
&=4\pi G \MX \f{D_\text{OL}D_\text{LS}}{D_\text{OS}}\,.
\end{align}
The differential probability of a beam of light being a lensed by a cosmological distribution of Macros is then
\begin{align}
d\t &= \nX\(z_L\) \s dt \notag\\
&=\f{3}{2} \Omega_\text{\tiny X} H_0^2 \(1+z_L\)^3 \f{D_\text{OL}D_\text{LS}}{D_\text{OS}} \(\f{dt}{d z_L}\) d z_L  \,,
\end{align}
where we have used $\nX(z_L) = \nX(0)\(1+z_L\)^3$, assumed the Macro population has not evolved significantly since $z_L$, used $\nX=\rX/\MX $ and then invoked the Friedmann equation to write the fractional density of Macros as $\Omega_\text{\tiny X}$. To the extent that the cross section is given by $\s$, this probability is therefore independent of $\MX$.

For a light signal traveling past a massive source, the total geodesic distance from the source to the observer is given by (\cite{Barnacka:2012bm})
\begin{align}
\Delta {\cal D} 
 \simeq  \f{1}{2} \f{D_\text{OS}}{D_\text{LS}D_\text{OL}} \(r_-^2 - r_+^2\) - 4 M G \ln{\f{r_+}{r_-}}\,,
\end{align}
where 
\begin{equation}
r_\pm = \f{1}{2}\(r_0 \pm \sqrt{r_0^2 + 4R_\text{\tiny E}^2}\)\,.
\end{equation}
This is a \emph{comoving} delay, however; in order to get the physical delay that accounts for cosmological expansion we note that the splitting of the image in time happens quite near the lens at redshift $z_L$, and so the physical delay is actually
\begin{equation}
\Delta r = (1+z_L) \Delta {\cal D}\,.
\end{equation}

For a typical lens redshift of $z_L\simeq 1$, the period (in energy space) of the modulation signal is
\begin{align}
E_p&=\f{2\p}{\abs{\Delta r}}\notag\\
&\sim
\f{10^{17}\text{g}}{\MX} ~~\text{MeV}
\end{align}
This should be compared with both the energy range and resolution of the telescope used for the observation, since in order to see a fringe pattern in the source's spectrum it is required that a phase of $\sim 1$ be visible, therefore both $E_p\lesssim E_{ran}$ and $E_p \gtrsim E_{res}$ is required (\cite{Barnacka:2012bm}).  Given characteristic values of $E_{ran}={\cal O}(1)~\MeV$ and $E_{res}={\cal O}(1)~\keV$ for gamma-ray telescopes, constraints are possible in the range $10^{17} \lesssim \MX \lesssim 10^{20}$ g.

Using this technique, the BATSE GRB data was used to rule out objects in the range $2\times10^{17} \lesssim \MX \lesssim 2\times 10^{20}$ g at $2\s$ from constituting a major fraction of the dark matter (\cite{Marani:1998sh}). Recently, the GRB data taken by the Fermi satellite was used to rule out $10^{17} \sim 10^{19.5}$ g at $2\s$ (\cite{Barnacka:2012bm}). The combined results allow us rule out Macros in the range $10^{17}\,\text{g} \leq \MX \leq 2\times 10^{20}\,\text{g}$.  The sources and lenses are at a characteristic distance of $d\sim $ Gpc, so equation \eqref{lensing_reqt} allows us to apply this constraint for $\sm\lesssim 1 \cmsg$. Lastly, we note that in \cite{Pani:2014rca} it is claimed that the constraints in \cite{Barnacka:2012bm} were obtained ignoring finite-source effects, so it is with this caution that we employ their results.

\sssec{Microlensing}

Microlensing of a background light source by an intermediate gravitationally-lensing object results in a short-term change in the observed brightness of the source as the lens passes near the line of sight. In order for the microlensing of a source to be seen, its angular size should be smaller than the Einstein ring of the lens, $\th_\text{\tiny E} = R_\text{\tiny E}/D_\text{OL}$, setting a minimum size of the ring at some characteristic distance. Taking a characteristic source radius of one solar radius ($R_\odot = 7 \times 10^5$ km), this requires
\begin{equation}
 \th_\text{\tiny E} \gtrsim \f{R_\odot}{D_\text{OS}}\,.
\end{equation}
For the Large Magellanic Cloud at a distance of about about $50 ~\kpc$, this amounts to $\MX \gtrsim10^{-7} M_\odot$. The transit time for an object to pass through the corresponding Einstein radius is roughly
\begin{align}
\Delta t = \f{2 R_\text{\tiny E}}{\vX}
\end{align}
which sets the range of possible constraints due to finite observing times. For example, assuming the source is in the Large Magellanic Cloud, a Macro of mass $\MX=10^{-7}M_\odot=2\times 10^{26}$ g would take about an hour to make the transit while $\MX=10\, M_\odot=2\times10^{34}$ g would take about a year.

Constraints were put on such massive objects through the monitoring of sources in the Small and Large Magellanic Clouds, ruling out dark matter candidates in the range $1.2\times 10^{26} - 6\times 10^{34}$ g  (\cite{Allsman:2000kg,Tisserand:2006zx, Carr:2009jm}).  Given characteristic distances of about $50$ kpc, equation \eqref{lensing_reqt} tells us we can apply that constraint when $\sm \lesssim 10^{-4} \cmsg$. In \cite{Griest:2013aaa}, observations provided by the Kepler satellite were similarly used to make constraints on the range $4\times10^{24} - 2\times 10^{26}\,$g. 
At typical distances of about 1 kpc, equation \eqref{lensing_reqt} indicates these limits are applicable for $\sm\lesssim 10^{-6} \cmsg$. 

In summary, microlensing tells us that
\begin{equation}
\MX \lesssim 4\times10^{24}\,\text{g}\,.
\end{equation}
These constraints are included in Figure \ref{model_independent}.



\sec{Model-Dependent Constraints}\label{model-dep_constraints}

\ssec{Electromagnetic Properties of Models I, II, and III}
There is an inherent difficulty in making predictions about the way  macro dark matter would behave electromagnetically without knowing its precise composition, even if it is Standard Model in origin.  Different models exist in the literature that predict wildly different properties, primarily owing to whether or not electrons (positrons) can penetrate into them. We can separate the nuclear-inspired Macro models into three broad classes: (I) the net core charge, $\QX=0$; 
(II) there is a non-trivial $\QX -\RX$ relationship, but the Macros do not admit $e^\pm$; (III) the same as II, but the Macros do admit $e^\pm$.

For model types II \& III we will assume the $\QX -\RX$ relationship is maintained by the requirement that the surface potential owing to the core (quark) charge remains constant, i.e.
\begin{equation}
V_0=\f{\QX\a}{\RX}\,,
\end{equation}
where $\a\simeq1/137$ and $V_0$ is fixed, implying
\begin{equation}\label{nuclear_Q-R_relation}
\abs{\QX}\simeq 1.1 \times 10^5\(\f{\abs{V_0}}{15~\MeV}\)\(\f{\RX}{10^{-8}\cm}\)\,.
\end{equation}
Model II makes for very tractable calculations, since the surface potential is fixed at a constant value, however model III requires additional statistical-mechanical considerations.

\sssec{Model III}
Since a detailed description of the core charge distribution within the Macro is unknown, we take it to be uniform. In the places/eras of interest, the Macros will be immersed in a fluid of protons, electrons and, depending on the era, positrons. We will assume the overall distribution of the fluid to be determined by the hydrostatic equilibrium between the fluid pressure and the electrostatic force.

Recall that the number densities and pressures of a fermion species are given by
\begin{align}
n_i &= \f{1}{\pi^2}\int_{m_i}^\infty d\tE \f{\tE \sqrt{\tE^2-m_i^2}}{e^{\(\tE+V_i-\m_i\)/T}+1}\\
P_i &= \f{1}{3\pi^2}\int_{m_i}^\infty d\tE \f{\(\tE^2-m_i^2\)^{3/2}}{e^{\(\tE+V_i-\m_i\)/T}+1}\,,
\end{align}
where here $\tE$ is only the relativistic part of the energy, i.e. $E=\tE+ V_i$, where $E$ is the full energy eigenvalue of the Hamiltonian. Using this definition, the chemical potentials, $\m_i$ are the same as their background values in the absence of the Macro. In the classical limit, applicable in the cases of interest, the exponential dominates in the denominators and one finds
\begin{align}
n_i&= e^{-V_i/T} \bar{n}_i\\
P_i&= e^{-V_i/T} \bar{n}_i T\,,
\end{align}
where the barred values are the background values, and we have defined $V_{e^{-}}\equiv V$ so that $V_{e^{+}}=V_{p}=-V$.

Since the system is taken to be spherically symmetric, the condition for electric hydrostatic equilibrium is
\begin{equation}\label{HydroEquil}
\f{d P}{d r} = \f{Q_\text{int}(r)\a}{r^2}\(-n_{e^-} + n_{e^+} + n_{p}\) \,,
\end{equation}
where $Q_\text{int}\(r\)$ is the (integer) charge within a sphere of radius, $r$, and is given by
\begin{equation}
Q_\text{int}\(r\)=
\begin{dcases}
\QX\(\f{r}{\RX}\)^3 + 4\p \int_0^r d\tr ~ \tr^2 \(n_p + n_{e^+} - n_{e^-}\)\\
\QX + 4\p \int_0^r d\tr ~ \tr^2 \(n_p + n_{e^+} - n_{e^-}\)\,,
\end{dcases}
\end{equation}
where the top and bottom lines apply for $r<\RX$ and $r>\RX$, respectively. In the limit $r\to0$, one can show that $dP/dr \to 0$ for a reasonably well-behaved potential\footnote{This is true if $V(r)$ is finite at $r=0$, for example.}, indicating
\begin{equation}
V'(0)=0\,.
\end{equation}
The other boundary condition on $V$ comes from the requirement
\begin{equation}
\lim_{r\to\infty} V(r) \to 0
\end{equation}
The value of $V(0)$ will generally not be known, except for when a full analytic solution for $V(r)$ is available, hence numerical ``shooting" will be required in order to numerically evolve $V(r)$ from $r=0$. The differential equation to solve is found by taking a derivate of \eqref{HydroEquil} with respect to $r$, which simply results in a version of the Poisson equation\footnote{This is consistent with our convention that $V=-e \phi$, where $\phi$ is the electrostatic potential; it satisfies $\del^2\phi = -\rho$, where $\rho$ is the charge density.}
\begin{equation}
V''(r)+\f{2}{r}V'(r)=4\p \a \bigg(\pm n\, \Theta(\RX-r) + n_p + n_{e^+} - n_{e^-}\bigg)\,,
\end{equation}
where $n\equiv3\abs{\QX}/(4\p \RX^3)$, and upper or lower signs refer to $\QX>0$ or $\QX<0$, respectively. After dividing by $T$, defining $q^2\equiv4\p \a n/T$, $y\equiv qr$, and $v(y)\equiv V/T$, we arrive at the dimensionless equation
\begin{equation}\label{dimless_v_eqn}
\ddot{v}(y)+\f{2}{y}\dot{v}(y)= \pm \Theta\(\yX-y\) + \f{n_p + n_{e^+} - n_{e^-}}{n}\,,
\end{equation}
where a dot is a derivative with respect to $y$ and $\yX\equiv q \RX$. Generally, this equation must be solved numerically; however, in the case $\abs{v(y)}\ll 1$ it is easy to show that 
\begin{equation}
n_p + n_{e^+} - n_{e^-} \simeq \(\bar{n}_p + \bar{n}_{e^+} + \bar{n}_{e^-}\)v(y)
\end{equation}
due to overall charge neutrality. Therefore the solution to \eqref{dimless_v_eqn} is approximately
\begin{equation}
v(y)=
\begin{dcases}
\mp \(\f{1}{a^2} - c_1 \f{e^{a y} - e^{-ay}}{y}\)~~~&\mbox{($y<\yX$)}\\
\pm c_2\f{e^{-ay}}{y}~~~&\mbox{($y>\yX$)}\,,
\end{dcases}
\end{equation}
where $a^2\equiv \bar{n}/n$ and $\bar{n}\equiv \bar{n}_p + \bar{n}_{e^+} + \bar{n}_{e^-}$. By the matching of $v(\yX)$ and $\dot{v}(\yX)$ at $\yX$ we learn
\begin{align}
c_1&=\f{1}{2a^3}e^{-a\yX}\(1+a\yX\)\\
c_2&=\f{1}{a^3}\(\sinh{a\yX} - a\yX \cosh{a\yX}   \)\,.
\end{align}

In order to use these solutions, we must establish their range of validity. Given that $\abs{v(y)}$ is known to decrease monotonically with increasing $y$, these solutions will be appropriate if $\abs{v(0)}\ll1$ and so we would like to know for what $\QX - \RX$ relation this is true. It is therefore checked in two limits:\R
\hphantom{OOOO}\bul~ $a\yX \ll 1$: Here $\abs{v(0)}\simeq \yX^2/2$, from which it is required that
\begin{equation}
\abs{\QX}\ll\f{2}{3}\f{\RX T}{\a}\,.
\end{equation}
\\
\hphantom{OOOO}\bul~ $a\yX \gg 1$: Here $\abs{v(0)}\simeq n/\bar{n}$, and therefore it is required that
\begin{equation}\label{4353452}
\abs{\QX} \ll \f{4\p}{3}\RX^3 \bar{n}\,.
\end{equation}
Given the above results, we consider two two eras/systems that are of interest: BBN and stellar cores.
\R
{\bf BBN ($T=1~\MeV$):} Here $\bar{n}\simeq 5 \times 10^{31} \cm^{-3}$, therefore $a\yX\simeq \f{\RX}{10^{-10}\cm}$. The mica limit indicated that such nuclear-dense Macros are only possible for $\RX\gtrsim 3\times 10^{-5}$ cm, and therefore the small $\abs{v}$ is guaranteed if \eqref{4353452} is satisfied. Given the $\QX -\RX$ relationship of \eqref{nuclear_Q-R_relation}, this requires
\begin{equation}\label{RX_limit_BBN}
\RX \gg 2\times 10^{-10}~\cm~ \sqrt{\abs{\f{V_0}{15~\MeV}}}\,,
\end{equation}
which is easily satisfied in nuclear-inspired models where $\abs{V_0}\sim {\cal O}(10)~\MeV$ (see e.g. \cite{Alcock:1986hz, Zhitnitsky:2006tu, Cumberbatch:2006bj}) \footnote{Although BBN occurs between temperatures of about 1 MeV to 0.1 MeV, it is during the higher temperature (earlier time) era that the majority of nucleon absorption would occur and have a possible effect.}.\\\\
{\bf Stellar Core ($T\simeq1~\keV$)}: Here we take the solar value of $\bar{n}=n_{\odot, \text{\tiny core}}\simeq 2 \times 10^{26} \cm^{-3}$, therefore $a\yX\simeq \f{\RX}{2\times 10^{-9}\cm}$. Again, the mica limit indicates that small $\abs{v}$ is guaranteed if \eqref{4353452} is satisfied. Given the nuclear model of \eqref{nuclear_Q-R_relation}, this requires
\begin{equation}\label{RX_limit_stellar}
\RX \gg 10^{-7}~\cm ~\sqrt{\abs{\f{V_0}{15~\MeV}}}\,,
\end{equation}
which is also easily satisfied in the representative models mentioned above. In summary, we have established that $V(0) < T$, and hence $V(\RX) < T$ in both of the above systems, indicating that there is no significant Coulomb barrier to prevent protons from entering the Marcos of model III during BBN or inside a typical stellar core today.


\ssec{BBN Limits on Model II}

A Macro could affect the path of standard model particles, possibly absorbing them or even catalyzing their decay, as in the scenario of supersymmetric Q-balls (\cite{Kusenko:2004yw}). If Macros absorbed a significant fraction of the ambient neutrons and a negligible fraction of protons, for example, the standard BBN predictions would be altered; this was noted,
 for example,  in the context of strange nuclear matter (\cite{Madsen:1985zx}). Since nearly all surviving neutrons during BBN end up in $^4$He, the primordial helium mass fraction, $X_4$ is
\begin{equation}
X_4=\f{2n_n}{n_n +n_p},
\end{equation}
which has been measured at the few percent level using observations of metal-poor extragalactic H II regions (\cite{Aver:2013wba}). A modest decrement in the relative abundance of neutron or protons would then significantly affect this observable quantity. Model types I and III would allow protons into the Macro at nearly the same rate as neutrons, so an effect on the primordial abundances of the light elements is not expected\footnote{Of course, this is true unless the Macros were to have swept up a large fraction of the total number of baryons, and this depends on $\sm$ and the Macro formation temperature. We would worry if
\begin{equation*}
\f{\Delta \MX}{\MX} \sim 10^6 \f{\sX}{\cm^2}\f{1\text{~g}}{\MX} T_{9,form}^{3/2} \sim {\cal O}(1)\,.
\end{equation*}
If the formation temperature was $150~\MeV$, for example, then $T_{9,form}\simeq 2\times 10^{3}$ and we would require $\sm \ll 10^{-10} \cmsg$ to ensure  $\f{\Delta \MX}{\MX}\ll 1$, thereby guaranteeing that most of the baryons were not swept up by the Macros. This would require $\MX\gg 1$ g, which is marginally satisfied by the mica constraints.}. \emph{The following constraints therefore only apply to model type II.}

Before comparing to theory 
one must be careful, however, as this effect is degenerate with other possibilities that could affect the measured value of $X_4$, such as the existence of extra relativistic species ($\Delta N_\n$) or errors in the measurement of the baryon fraction ($\Omega_\text{\tiny b}$), loosening any constraint to some degree; the current experimental error bars in e.g. \cite{Aver:2013wba} are still large enough to justify neglecting this degeneracy, however. The results from that work indicate that the observed abundance $X_4^\text{obs}\simeq 0.25 \pm 0.01$, while the standard theoretical prediction, $X_4^\text{std}=0.2485 \pm 0.0002$. Therefore if we characterize a deviation due to Macros by 
\begin{equation}
X_4=X_4^\text{std}+\Delta X_4^\text{Macro}
\end{equation}
we conclude the observational bound
\begin{align}\label{Delta_X_4_bound}
-0.01 \lesssim \Delta X_4^\text{Macro} \lesssim 0.01 
\end{align}
since the uncertainty on $X_4^\text{std}$ is negligible.
 
To calculate $\Delta X_4^\text{Macro}$, we use the comoving proton and neutron number densities
\begin{align}
{\cal N}_{n, p} &\equiv a(t)^3~n_{n, p}\,, 
\end{align}
where $a(t)$ is the cosmological scale factor. For the accuracy required here it suffices to use the approximate evolution equations
\begin{align}
\dot{{\cal N}_n} &= -\(\Gamma_n + \Gamma_{nX} \){\cal N}_n\label{Y_n_evol}\\
\dot{{\cal N}_p} &=  +\Gamma_n {\cal N}_n  - \Gamma_{pX}{\cal N}_p\,.\label{Y_p_evol}
\end{align}
Here $\Ga_n$ is the neutron decay rate, and $\Ga_{nX}$ ($\Ga_{pX}$) is the rate of neutron (proton) absorption\footnote{This is also appropriate if the Macro catalyzes the baryon to decay to something non-baryonic.} by Macro. Most relevant for computing the effects on the $^4$He abundance is the proton to neutron ratio
\begin{equation}
\a(t)\equiv \f{n_p}{n_n} = \f{{\cal N}_p}{{\cal N}_n}\,,
\end{equation}
from which we may write
\begin{equation}
X_4(t)=\f{2}{\a(t) + 1}\,.
\end{equation}
From \eqref{Y_n_evol} and \eqref{Y_p_evol}, $\a(t)$ obeys the evolution equation
\begin{equation}
\dot{\a}(t)= \Ga_n + \(\Ga_n +\Ga_X\)\a(t)\,,
\end{equation}
where $\Ga_X\equiv\Ga_{nX} - \Ga_{pX}$.  The solution is
\begin{equation}
\a\(t\) = \(\a_0+ \int^t_{t_0} d\tilde{t} ~\Ga_ne^{-\int^{\tilde{t}}_{t_0} dt'~ \(\Ga_n + \Ga_X\) }  \)e^{\int^t_{t_0} dt''~\(\Ga_n + \Ga_X\)} \,,
\end{equation}
where $\a_0$ is set by the proton-to-neutron ratio at the time of weak-interaction freeze-out, which we presume to be unaffected by the presence of Macros. Technically, $\Ga_n$ is temperature dependent (\cite{Alpher:1953zz, Dicus:1982bz}) but it was found that including this effect only changes our results by roughly $0.1$ per cent. The Macros' effect would apparently be small, so we expand in $\int  \Ga_X$, finding 
\begin{equation}
\a(t_B)=\a^\text{std}(t_B)\(1+a\)-b,
\end{equation}
where $\a^\text{std}(t_B)$ is the standard value and
\begin{align}
a&=\int_{t_F}^{t_B}dt~\Ga_X\\
b&=  e^{\int^{t_B}_{t_F}dt~\Ga_n} \int_{t_F}^{t_B}d\tilde{t}~\Ga_n~e^{-\int_{t_F}^{\tilde{t}}dt~\Ga_n} \int_{t_F}^{\tilde{t}}dt~\Ga_X\,.
\end{align}
We denote $t_F$ as the time of weak-interaction freeze-out and $t_B$ as the time of the deuterium bottleneck breaking; to good approximation, this defines the time of efficient $^4$He production -- thus $X_4$ is largely determined by $\a(t_B)$.  Given that $X^\text{std}\simeq 0.25$ (or $\a^\text{std}(t_B)\simeq7$), we find
\begin{equation}\label{Delta_X_4_Macro}
\Delta X_4^\text{Macro}\simeq -\f{7a-b}{32}\,.
\end{equation}

To perform the integrals in $a$ and $b$, we change our integration variable to temperature using the time-temperature relation
\begin{equation}
t=\f{\th}{T_9^2} ~\text{s}\,,
\end{equation}
where $T_9$ is the temperature defined in units of $10^9$K and $\theta$ depends on the number of relativistic degrees of freedom. As in \cite{esmailzadeh}, we find
\begin{equation}
\th=
\begin{dcases}
&99.4,~~~~ T_9>5\\
&178, ~~~~~T_9<1\,,
\end{dcases}
\end{equation}
assuming the standard value of $N_\text{eff}=3.046$. In what follows, we use the values\footnote{These were the values used in \cite{esmailzadeh}. Changing them by 10 per cent affects the integrals only at the few-percent level.} $T_{9,F}=9.1$ and $T_{9,B}=1$
and numerically determine $a$ and $b$ for different Macro properties. Integrating through this range requires an interpolation of $\th(T_9)$ in the region $1\leq T_9 \leq 5$; to do this we choose a hyperbolic tangent centered around $T_9=2$: 
\begin{equation}\label{theta_interpolation}
\th(T_9) \simeq \th_\text{max} - \f{1}{2}\(\th_\text{max} - \th_\text{min} \)\(\tanh{\[T_9-2\]} +1\)\,, 
\end{equation}
where $\th_\text{max}=178, \th_\text{min}=99.4$. The general formula \eqref{theta_interpolation} is sufficient to match all of the values quoted in Table 15.5 of \cite{Weinberg} to an error of less than 10 per cent, which is sufficient for our purposes\footnote{\cite{Weinberg} used two neutrino species to calculate Table 15.5 therein; to compare to it, $\th_\text{max}$ and $\th_\text{min}$ must be corrected to account for this smaller number of relativistic fermions.}.

Since a neutron is neutral, its absorption rate by Macros is given by
\begin{align}
\Gamma_{nX}&=\<\f{\rX}{\MX} \sX v\>\\
&=5.1 \times 10^4 \times T_9^{7/2} \f{\sX}{\MX} \f{\text{g}}{\text{cm}^2 \text{ s}}\,,
\end{align}
where we have used the thermally-averaged neutron velocity $v_n=\sqrt{8T/(\pi m_n)}$ and inserted $\rX =3H_0^2/(8 \pi G)\Omega_c\(T/T_0\)^3=0.93\times 10^{-3} \Omega_c h^2 T_9^3 ~\gcmc$ with the Planck value of $\Omega_c h^2=0.1199$ (\cite{Ade:2013zuv}).


For a proton of energy $E_p$ incident on a Macro with surface potential $V(\RX)$ the effective cross-section, $\s_\text{\tiny X,eff}=\sX\(1-\f{V(\RX)}{E_p}\)$. This must be thermally averaged along with the velocity, resulting in
\begin{equation}
\<\s_\text{\tiny X,eff}v\>=\sX\times
\begin{dcases}
e^{-\f{V(\RX)}{T}}\<v\> , ~~~~~~~~~~~&V(\RX)\geq 0\\
\(1-\f{V(\RX)}{T}\)\<v\>,& V(\RX)< 0\,.
\end{dcases}
\end{equation}
It then follows that the proton absorption rate may written in terms of the neutron rate as
\begin{equation}
\Gamma_{pX}=\Gamma_{nX}\times
\begin{dcases}
 e^{-V(\RX)/T}, ~~~~~~~~~~~&V(\RX)\geq 0\\
\(1-\f{V(\RX)}{T}\),& V(\RX)< 0\,.
\end{dcases}
\end{equation}
For positive surface potentials there is a Boltzmann-Coulomb suppression for the absorption rate, whereas if it is negative there is a Coulomb enhancement. We do not allow $V(\RX)$ to evolve, in accordance with nuclear model II.
 
\begin{figure}
  \begin{center}
    \includegraphics[scale=.45]{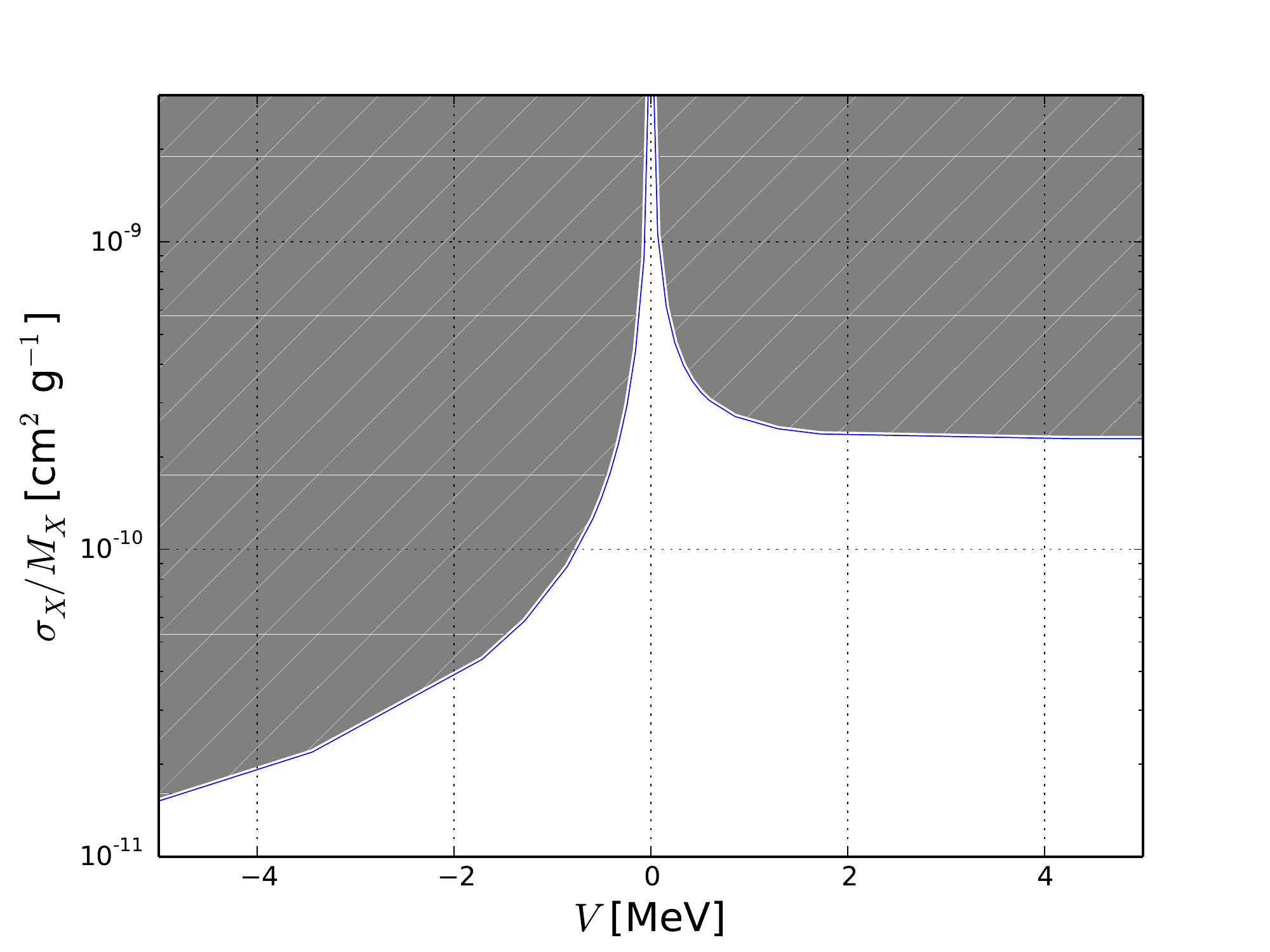}
  \end{center}
\caption{Model-dependent constraints from BBN on the reduced cross section as a function of the Macro surface potential, in model II. The parameter space ruled out lies in the grey region.}
\label{BBN-interp}
\end{figure}

Combining the predicted Macro effect of $\Delta X_4^\text{Macro}$, equation \eqref{Delta_X_4_Macro} and the observational bound, equation \eqref{Delta_X_4_bound}, we find that
\begin{equation}
\f{\sX}{\MX} \lesssim 8\times10^{-11}      \abs{\f{V(\RX)}{\text{MeV}}}^{-1} \cmsg \,,
\end{equation}
for $V(\RX)\lesssim 0.01$ MeV, while for $V(\RX)\gtrsim 1$ MeV we find the bound asymptotes to
\begin{equation}
\f{\sX}{\MX} \lesssim 2 \times 10^{-10} \cmsg\,.
\end{equation}
In between there is a transition, illustrated in Figure \ref{BBN-interp}.

As an example, for nuclear-type models with $V(\RX)\simeq-20$ MeV as suggested in e.g. \cite{Alcock:1986hz, Zhitnitsky:2006tu, Cumberbatch:2006bj}, the constraint would be $\s/\MX\lesssim 4 \times 10^{-12}\cmsg$ if such Macros absorb protons. At nuclear density ($\simeq 3.6\times 10^{14}~\gcmc$) this translates to a bound on the mass of $\MX\gtrsim 3\times 10^5$ g, an improvement over the limit from ancient mica by nearly 4 orders of magnitude.


%
%
%

\ssec{``Converting" Dark Matter}

Should the dark matter be of some more stable form of matter than ordinary baryons, it is conceivable that it could convert astrophysical objects to its form; this is thought to occur in various models of stable strange matter, where normal baryonic matter is converted to this state (see e.g.  \cite{Witten:1984rs} or \cite{Alcock:1986hz} ). If the dark matter were of this type of ``converting" variety, it could potentially convert any target that it gets captured inside of, such as the Sun or a neutron star, for example. This is necessarily model dependent -- in model II, for example, the Macro cannot easily absorb protons when $V_0\gg1 ~\keV$ and so it would be incapable of converting a typical star. On the other hand, a neutron star can be used, in principle, for constraints in a more model-independent way since the absorption of neutrons is independent of the Macro charge.


For a Macro of mass $\MX$ passing through an astrophysical spherical target on a secant line of length, $D$, it can be shown that its velocity in the target evolves according to
\begin{equation}\label{dvdx}
\f{dv}{dx}= -\r_\text{\tiny T} \f{\sX}{\MX(x)} v(x) + \f{4\p}{3} \f{G \r_\text{\tiny T} \(\f{D}{2}-x\)}{v(x)}\,,
\end{equation}
where $\r_\text{\tiny T}$ is the average target density. $\MX(x)$ is constant for elastically-scattering Macros and is a linear function of $x$, the passage depth, in the inelastic case. We define the dimensionless parameter $\a\equiv\r_\text{\tiny T} \sX R_\text{\tiny T}/\MX$, which gives a measure of the ability of the target to capture a Macro. Equation \eqref{dvdx} is soluble in both the elastic and inelastic cases, however it suffices to note that the solutions are identical to ${\cal O}(\a)$ and therefore, upon the Macro's exit, the final velocity is given by
\begin{equation}\label{v_f_eqn}
v_f^2 = v^2_0 - \f{D}{R_\text{\tiny T}} \(2 v_0^2 +\f{1}{6} \(\f{D}{R_\text{\tiny T}}\)^2 v_e^2\) \a + {\cal O}(\a^2)\,.
\end{equation}

\sssec{Orbital Capture Requirements}

Because of spherical symmetry, it should be the case that the average value, $\<D\>=4/3 R_\text{\tiny T}$, which we will simply insert everywhere  $D$ appears in \eqref{v_f_eqn}. The orbital capture requirement of $v_f < v_e$ can then be written as 
\begin{equation}\label{1st_orbital_cap_reqt}
 v^2_0 - \f{8}{3}\(v_0^2 +\f{4}{27} v_e^2\) \f{\r_\text{\tiny T} \sX R_\text{\tiny T}}{\MX} < v_e^2
\end{equation}
from which it follows that
\begin{align}\label{orbital_cap_reqt}
\f{\sX}{\MX} 
\gtrsim 5\times 10^{-13}\cmsg~ \(\f{R_\text{\tiny T}}{R_\odot}\)^3\(\f{\vX}{250\text{~km s$^{-1}$}}\)^2
\end{align}
is needed for orbital capture\footnote{We have used the relation $v_0^2 = v_e^2 + \vX^2$, following from energy conservation, where $\vX$ is the asymptotic velocity of the incoming dark matter and satisfies $\vX < v_e$ (or $\vX \ll v_e$) in the astrophysical systems of interest here.}. The Macro may then be considered internally captured as the time scale for that process is significantly smaller than the 5 Gyr age of the Sun.

\sssec{Constraints on Nuclear-Dense Macros}

Specializing our discussion to the nuclear case, the Macro reduced cross section follows $\sX/\MX=2.4 \times 10^{-10} \(1~\text{g}/\MX\)^{1/3}\cmsg$. Equation \eqref{orbital_cap_reqt} indicates that Macros with masses $\gtrsim 10^5$g would not typically be captured in the Sun. The dark matter presumably follows a velocity distribution, however, so there will be some fraction of Macros whose $\vX$ is less than some critical value, $\tilde{v}$, required for capture. From \eqref{1st_orbital_cap_reqt}, we find this velocity is approximately
\begin{equation}\label{vt_eqn}
\tilde{v}\simeq 5~\text{km s$^{-1}$}~ \(\f{R_\odot}{R_\text{\tiny T}}\)^{3/2} \(\f{10^{18}\text{~g}}{\MX}\)^{1/6}
\end{equation}

We will assume that the dark matter follows a Maxwell-Boltzmann velocity distribution, i.e. the probability distribution function is
\begin{equation}
f(v) = \sqrt{\f{2}{\p}} \f{v^2}{\s_v^3}e^{-\f{1}{2}\f{v^2}{\s_v^2}}\,,
\end{equation}
where $\s_v= \sqrt{\p/2} \<v\>/2$ and the distribution is normalized to satisfy\footnote{Technically the distribution would need to be truncated at the escape velocity of the galaxy, $v_\text{esc}$. This would affect the overall normalization, however, models indicate that $v_\text{esc}\gtrsim 2 \<\vX\>$ (see e.g. \cite{Fairbairn:2012zs}), therefore extending the integral to infinity introduces an error no larger than a few percent.}
\begin{equation}
1=\int_0^\infty dv~f(v)\,.
\end{equation}
The probability of a Macro having an asymptotic velocity less than this is
\begin{align}\label{low_vt_probability}
P(\vX < \tilde{v})&=\int_0^{\tilde{v}} d\vX ~f(\vX)\notag\\
& \simeq \f{32}{3\p^2}\(\f{\tilde{v}}{\<\vX\>}\)^3\,,
\end{align}
where the second line applies in the limit $\tilde{v}\ll \<\vX\>$.
\newpage
~\\
{\bf Solar Constraint}\\
\\
It is clear that nuclear-dense macros of low mass would have no trouble being captured by the Sun, so we focus our attention on the high-mass regime where $\tilde{v}$ is very small compared to $\<\vX\>$. The total number of captures, $N_\text{cap}$, is the product of the probability given in \eqref{low_vt_probability} and the total number of passages through the Sun in its lifetime of 5 Gyr as determined from the rate given in \eqref{impact_rate_2}.  The value of $N_\text{cap}$ for the Sun apparently decreases faster than $\<\vX\>^{-2}$ with increasing $\<\vX\>$; due to a large uncertainty on the correct value of $\<\vX\>$ (see e.g. \cite{Fairbairn:2012zs}) we will conservatively use the rather large value of $\<\vX\>=300$ km s$^{-1}$ in what follows below.  From \eqref{impact_rate_2} we then find
\begin{equation}
N_\text{pass}\simeq 3 \times 10^5 \(\f{10^{18}\text{g}}{\MX}\)\,.
\end{equation}
Inserting \eqref{vt_eqn} into \eqref{low_vt_probability}, we find that
\begin{equation}
P(\vX < \tilde{v})\simeq 5\times10^{-6} \(\f{10^{18}\text{~g}}{\MX}\)^{1/2}
\end{equation}
and therefore
\begin{equation}
N_\text{cap}\simeq 2\times\(\f{ 10^{18}\text{~g}}{\MX}\)^{3/2}\,.
\end{equation}

The fact that our Sun still shines and is well described using ordinary baryonic physics indicates it has not been converted. Using its existence to make a constraint necessarily introduces an anthropic selection bias, so we can only make the following statistical statement. Assuming the conversion (or destruction, for that matter) of our Sun to be a random process, its survival probability in this context is given by
\begin{equation}
P_\text{survival}=e^{-N_\text{cap}}\,.
\end{equation}
Therefore we can, for example, conclude a bound of
\begin{equation}
\MX \gtrsim 7\times10^{17}\,\text{g}
\end{equation}
at $95$ per cent confidence. It becomes exponentially unlikely that such converting Macros exist at lower masses. We therefore rule out converting, nuclear-type Macros with the electromagnetic properties of model I, model II for $V_0 \lesssim 1~ \keV$, and model III in this mass range.\\
\R
{\bf Neutron Star Constraint}\\
\\
Here, the orbital capture requirements are easily satisfied by any nuclear-dense Macro in the mass range of interest. Therefore $N_\text{cap}$ is determined solely by the number of passages through a typical neutron star  -- for this reason we again conservatively use $\<\vX\>=300$ km s$^{-1}$ and find
\begin{equation}
N_\text{cap}\simeq 3 \times \(\f{10^{18}\text{g}}{\MX}\)\(\f{T_\text{NS}}{5\text{~Gyr}}\)\,,
\end{equation}
where $T_\text{NS}$ is the age of the neutron star and we have assumed that the dark matter density around it to be comparable to that of our local galactic neighborhood. 
The use of neutron stars to make a constraint on converting dark matter offers its own difficulty as it requires clear observational differences between an ordinary neutron star and whatever the converted object would be, e.g. a star that is composed, at least in part, of strange nuclear matter. For example, a pulsar's composition affects its seismology (\cite{Madsen:1998qb}); this in turn affects its spin-down rate through an alteration in its gravitational wave emission, and this has the potential for observation through pulsar timing (\cite{Alford:2013pma}).


The crust of a neutron star, which also plays a vital role in the modeling of pulsar glitches, would determine the nature of the Macro capture. It is thought to be a dense material consisting of positively-charged nuclei, with a column density of perhaps $4\times10^{15}$g/cm$^2$; this implies that nuclear-dense Macros with masses less than $\sim 10^{18}$g might only get captured in the crust, unable to penetrate to the deeper neutron-abundant regions (\cite{Madsen:1989pg}).  It therefore appears that the constraints possible from neutron stars are only complementary to those from stellar objects.

\sec{Fruitless Ideas}\label{ideas}

~\\
{\bf Macro Luminosity}: Because of the mass and size of the range of Macro parameters, they are not expected to significantly heat up material or be heated by collisions with baryonic matter in the galaxy. Regardless of their formation mechanism, we expect that they have become (and will remain) cold, and therefore dark in typical astrophysical systems.
 To justify this claim, we approximate the luminosity of a Macro at temperature $T_\text{\tiny X}$, to be given the Stefan-Boltzmann relation 
\begin{equation}\label{energy_loss}
{\cal L}_X= \f{\p^3}{15} \RX^2 T_\text{\tiny X}^4\,.
\end{equation}
At best, the Macro could hold a constant temperature by maintaining a balance between energy absorption and radiation. The rate at which energy is acquired from a surrounding gaseous environment is, at most, given by\footnote{The relevant velocity at which energetic particles impact the Macro is $\vX$, not the much smaller gas velocity.}
\begin{equation}\label{energy_acq}
\p \RX^2 n_{gas} T_{gas} \vX
\end{equation}
and therefore, by equating \eqref{energy_loss} with \eqref{energy_acq}, the Macro temperature is expected to be
\begin{equation}
T_\text{\tiny X}\lesssim  \sqrt[4]{\f{15}{\p^2} n_{gas} T_{gas} \vX}\,,
\end{equation}
which is independent of the size of the Macro, so long as it is macroscopically large. With $\vX \approx 10^{-3}c$, we consider a few different systems: for molecular clouds in the interstellar medium, $n_{gas}\approx 10^6~\text{cm}^{-3}$ and $T_{gas}\approx 10\,$K indicating  $T_\text{\tiny X}\approx 5\,$K; in the warm ionized medium $n_{gas}\approx 1~\text{cm}^{-3}$ and $T_{gas}\approx 8000\,$K, resulting in $T_\text{\tiny X}\approx 1\,$K; the intracluster medium can have gas temperatures as high as $T_{gas}\approx 10^{8}\,$K, but $n_{gas} \approx 10^{-3}~\text{cm}^{-3}$ so $T_\text{\tiny X}\approx 2\,$K. Presumably, the Macro temperature would not drop below the CMB temperature of $2.7\,$K, however.
\R
{\bf Early universe neutrinos}: Compared to the canonical cosmological model where less than 20 per cent of the matter is baryonic, the actual total number of baryons could be perhaps a factor of 6 larger if we live in a universe where the dark matter consists of Macros made of quarks. Of course, this depends on the model and how $\MX$ scales with the baryon number, $\BX$; for example, $\MX\propto \BX^{8/9}$ in \cite{Zhitnitsky:2002nr}. A consequence of this is that, assuming that the total baryon-minus-lepton number ($B\!-\!L$) remains fixed,  the total lepton number would be commensurately larger than in the canonical case. However, since the neutrino density is almost entirely determined by thermal considerations and is already on the order of $10^{9}$ larger than the baryon density, this appears to be a negligible effect. The neutrino mean free path would be also affected at the ${\cal O }(1\!-\!10)$ level\footnote{The Macros would not volume shield in the mass range of interest if they were nuclear-dense objects made of quarks.}, however, this isn't obviously observable as the neutrino thermal history remains unchanged.
\R
{\bf CMB Opacity}: One might wonder if the CMB opacity would be affected by 
a difference in the electron number density, $\Delta n_e$, that results from the presence of positively charged baryonic Macros. The change in electron number density depends on the Macro charge, scaling as $\Delta n_e \sim \QX/\BX$ if $\MX\propto \BX$. In any reasonable model, if $\QX$ scales with $\BX$ it will do so as $\BX^p$, with $p\leq2/3$ for energetic reasons. Since terrestrial constraints indicate $\MX\gtrsim1$ g, or $\BX\gtrsim10^{24}$, this strongly suggests that $\Delta n_e$ would be negligible.

\sec{Conclusions}\label{conclusions}

The nature of dark matter is still largely unknown. For this reason,  it is prudent to hedge our bets on what it might be, keeping an open mind and focusing on what the observational constraints actually are -- in particular for objects that interact strongly with themselves and ordinary matter, and could plausibly be accounted for within the Standard Model. Here, we have considered the class of strongly interacting dark matter, which we call \emph{Macros}, that would have macroscopic size and mass. We have illuminated the constraints on regions of the geometric cross section vs. mass parameter space ($\sX\!-\!\MX$) between about $10^{-15} - 10^{33}~\cm^2$ and $10^{-12} - 10^{34}$ g, assuming the Macros have a single mass.

Ancient mica samples, the CMB and large-scale structure, as well as various gravitational lensing observations constrain only a portion of the above-mentioned parameter space. Likewise, the reduced cross section ($\sm$) can be constrained as a function of Macro surface potential in a certain class of models wherein Macros are capable of removing baryons from the standard primordial nucleosynthesis process.  Rather large regions of parameter space remain unconstrained, notably for nuclear-dense Macros of masses between $55 - 10^{17}$ g and $2\times10^{20} - 4\times10^{24}$ g, assuming Macros do not destabilize ordinary matter. If they do, the first window is entirely closed because of its effect on the sun.

It is conceivable that other observations not considered here can be used to make marginal improvements on the Macro constraints at low mass. Beyond the mica limits ($\MX\gtrsim 10^2$ g), however, the Macro flux would drop below $10^{-2} ~\text{km}^{-2}~\text{yr}^{-1}$ and Earth-based observations are ever more limited. It is also of note that, in the unconstrained range of Macros masses between $10^2 - 10^{17}$ g, there would be between 0.1 and $10^{14}$ Macros occupying the sphere enclosed by the Earth's orbital radius at any given time. It might be possible in the future to probe this region through local observations in our solar neighborhood. It may also be significant that, in this scenario of high-massed constituents, the dark matter's approximation as a fluid breaks down at much larger scales than in the standard WIMP scenario. This and the possibility for dark matter to interact strongly with baryons may have interesting (and observable) astrophysical consequences.

As this manuscript was being prepared the preprint by \cite{Burdin:2014xma} came to our attention.
\R
The authors thank Claudia de Rham, Andrew Tolley, Craig Copi, Tom Shutt, Dan Akerib, Adam Christopherson, Donnino Anderhalden, Dan Snowden-Ifft, and Amanda Weltman for discussions. We are particularly grateful to Celine Boehm for many discussions about the  collisional damping effect and how this can be used to constrain Macro-photon interactions. We would also like to thank the CERN Theory group for their hospitality and support during the initial phases of this work. During this work two of the authors (DMJ and GDS) were supported by Department of Energy grant DOE-SC0009946. One of the authors (DMJ) would also like to acknowledge support from the Claude Leon Foundation. 

\bibliographystyle{mn2e}
\bibliography{MacroDM_bib}

\end{document}